# Towards a "core" genome: pairwise similarity searches on interspecific genomic data.

Bradly J. Alicea[1], Marcela A. Carvallo-Pinto[2], Jorge L.M. Rodrigues[3]

[1] Department of Telecommunication, Information Studies, and Media, Michigan State University, East Lansing, MI  [2] Department of Plant Biology, Michigan State University, East Lansing, MI  [3] Center for Microbial Ecology, Michigan State University, East Lansing, MI.

## Abstract

The phenomenon of gene conservation is an interesting evolutionary problem related to speciation and adaptation. Conserved genes are acted upon in evolution in a way that preserves their function despite other structural and functional changes going on around them. The recent availability of whole-genomic data from closely related species allows us to test the hypothesis that a "core" genome present in a hypothetical common ancestor is inherited by all sister taxa. Furthermore, this "core" genome should serve essential functions such as genetic regulation and cellular repair. Whole-genome sequences from three strains of bacteria (*Shewanella* sp.) were used in this analysis. The open reading frames (ORFs) for each identified and putative gene were used for each genome. Reciprocal Blast searches were conducted on all three genomes, which distilled a list of thousands of genes to 68 genes that were identical across taxa. Based on functional annotation, these genes were identified as housekeeping genes, which confirmed the original hypothesis. This method could be used in eukaryotes as well, in particular the relationship between humans, chimps, and macaques.

## Introduction

The study of gene conservation is an intriguing scientific problem that has consequences for understanding functional differences between species and biological variation in general. The best starting point for getting a handle on what elements of the genotype are shared across species is to look at model organisms in which this phenomenon is very simple. Three strains of the marine bacterial species *Shewanella* sp. provide such a model that features a relatively small number of genes, a tightly packed genome that consists mainly of coding regions (Mira et al. 2001), and a large proportion of genes for which function is known.

Across animals, plants, and bacteria, a wealth of 532 whole-genome sequences is now available, and this number is expected to increase to 4,000 by 2010 (Overbeek et al. 2005).  Although much has being gained from analyzing a single genome from its genetic potential standpoint, there were only few attempts for genome cross comparisons among different taxa. Whole genome sequence comparison allows for identification of higher order traits to be shared between species such as codon usage, conservation of gene order (e.g. synteny), regulatory mechanisms, and gene dataset conservation (Nierman et al. 2004).

Gene conservation, even among very divergent species, suggests that these genes are under strong evolutionary constrains. Bejerano et.al (2004) have used statistical tests for natural selection to show that ultraconserved genes undergo strong positive selection in their sequence composition so that coding regions can serve the same functions across taxa These conserved set of genes are so important that any modification could harm life itself, resulting in species extinction. Hence, the set of genes maintained across species is a good historical record of divergence from a common ancestor.

Evolutionary inference between species can only be made through phylogenetic reconstruction. The accuracy of the reconstruction can be jeopardized due to species sampling biases, the genetic signatures of orthologous and paralagous genes, and possible horizontal gene transfer (Cicarelli et al. 2006). Because the evolution of a single gene may not reflect the evolution of a species (different rates of evolution), recent analysis has suggested the possibility of extracting a coherent phylogenetic pattern using a core set of genes to be found in all species (Cicarelli et al. 2006, Daubin et al. 2003).

Extracting a set of core genes shared among all species seems an inherently difficult task. The difficulty extends from different genome sizes and number of genes to the selection of percent identity cut-off values to be used in pairwise interspecific comparisons. Owing to these difficulties, few studies have attempted to answer these questions (Konstantinidis and Tiedje, 2005a and 2005b). However, the state of computational and sequencing technology is now ripe for a broader application of this method.

In this study, we report the development of an extensible procedure to identify conserved genes among different species. A conservative set of criterion was used when selecting the gene core set by taken into account only true gene orthologs and minimizing the risk of utilizing hidden paralogous evolving at different evolutionary rates. The degree of gene conservation is user specific and can be used for closely related as well as widely diverse species. This approach can be extended to any set of taxa as more plant and animal genomes become available.

## Materials and Methods

Reciprocal Blast searches were set up for the following species pairs: MR-4 (A) and MR-7 (B), MR-7 (A) and MR-1 (B), and MR-1 (A) and MR-4 (B). All gene sequences for all taxa were taken from GenBank (Species names and accession numbers: *Shewanella* sp. MR-1, AE014299; *Shewanella* sp. MR-4, CP000446; and *Shewanella* sp. MR-7, CP000444) for each species of *Shewanella*. All datasets were functionally annotated and converted into Fasta format by Artemis Version 9 (Sanger Institute, Wellcome Trust, Cambridge, UK).

**Creating a "toy" problem**

The first goal was to create a three-taxon "toy" problem, that is, one that could be both solved easily and be evolutionarily informative. The procedure was conducted in three steps (see Figure 1). The first step involved taking all of the genes from species A

(in each pair) and using them as query in a blast search against all the genes from species B. This results in 3 lists called "genes A" vs. "genes B" (one for each pair of species). The second step was to filter these lists of matches using an E-value lower than $1e^{-10}$. This results in a limited list of matches which were considered for further analysis. The third step involved conducting a reciprocal blast search by taking all sequences from each list generated in step two and using them as query sequences in an additional blast search against all the genes from species A (for each pair).

**Blast Searches**

All Blast searches were conducted on the HPCC (high-performance computing cluster) server at Michigan State University. A script similar to the Table 1 was used to submit and conduct the Blast search on HPC. This script defines the basic structure of the forward search. There are several aspects of the HPC job submission script that one must be aware of before running a search[1]. Using these parameters (and assuming minimal queue time), each job was finished in a little under 30 minutes. In addition, several flags were defined on the blast command line (lines 4 through 8 in Figure 1). These were defined in a way that allowed us to conduct the reciprocal search with minimal effort. A number of specific values were used to define the flags in our search. One flag value is of particular interest; the -b flag was set to a value of 3 to prevent multiple results involving the same gene from either the query and reference sequences. While by no means foolproof, this was done to avoid paralogous matches as much as possible. Additional visual inspection of the data was done post-hoc to filter any other duplicate matches.

**Formatting the datasets**

Table 2 shows the code used for formatting the DNA and protein sequence .fasta files into searchable databases[2]. Nucleotide sequences were tried first to confirm the experimental design, while protein sequences were used for the analyses reported here.

**Installing Blast 2.2.16 on the server**

To run a Blast search on an HPC cluster, a copy of the UNIX binary files must be installed in a user's personal scratch space. The command line in the submission script must point to the install location (to all relevant child destinations of the home directory) every time a program associated with Blast is run (see Table 2 for example). For additional details of this process, see Appendix.

**Details on running the job**

Blast (NCBI; NIH, Bethesda, MD) version 2.2.16 was used to generate these searches. The searches were conducted on the High Performance Computing Center (HPC) cluster (http://www.hpcc.msu.edu or ftp://hpc.msu.edu, port 22) at Michigan State University. A total of six job submission files were created to run the searches, and each file represented a specific job run on the cluster (see Table 1 for their structure)[3].

---

[1] Every job submitted to Blast must have several lines of #PBS commands. Most importantly, a walltime and the amount of computational resources to be used on a particular job must be defined. In Table 1, a walltime of 8:0:0 (8 hours) and resources of 1GB were specified.

[2] This code generates one .psq, one .phr, and one .nin or .pin file as output per input .fasta file. The flag -p defines whether the input .fasta file contains nucleotides (-p F) or proteins (-p T). If -p F is used, a .nin file is created, whereas if -p T is used, a .pin file is created.

**Deriving the core genome**

Generating the "core" list involves taking the reciprocal best matches (derived from the reference sequences used in the reciprocal search) with a very high degree of similarity from each group and comparing them with each other. To filter all results with a 90 percent identity or above, the Python script shown in Table 3 was used. Reciprocal searches for all three species pairs yielded three independent lists. Each list represented best matches and their associated descriptive statistics. A list of common genes was then generated that included only those genes that were common to all three Blast-generated lists. The following procedure was used: two lists were compared first, and then the matches from those lists were compared with the third. Descriptive data (i.e. sequences and functional information) was then retrieved from the original .fasta files to determine the characteristics of this "core" list.

## Results and Discussion

In this work we have compared the genomes of three different bacteria, two strains of *Shewanella* sp. MR4 and MR7, and *Shewanella oneidensis* MR1. In order to obtain a common set of genes between each pair of genomes, we needed to identify orthologous genes between them. For this, we made the following assumption: the best reciprocal match between two genomes gives us orthologous genes. In addition, we used the following criteria for best match: 90 percent of identity between the pair of proteins and E value of 1E-10 based on the fact that E value represents the number of times this match would be expected to occur by chance, and we think that 1E-10 it is a very low likelihood that this will be the case. Konstantinidis and Tiedje (2005) did a similar type of study on another portion of the *Shewanella* clade using this threshold with positive results.

Given these considerations, we obtained 988 genes in common between MR4 and MR7, 881 genes between MR1 and MR4, and 619 genes between MR1 and MR7 (Figure 1). Once we obtained the common set of genes for each pair, we looked at the number of genes in common in these 3 lists, the "core set of genes". There were 68 proteins in common. In addition, a visual survey of these datasets revealed that there were changes in syntenic structure (see Wei et al. 2002) between genomes. The first set of reciprocal Blast searches indeed confirmed that large-scale rearrangements within the *Shewanella* genome have occurred during evolution.

---

[3] A job can be submitted to the cluster by typing the following command: qsub yourjob.sh, where yourjob is the name of a specific .sh file. After a job is submitted, the server responds with a job number (usually six digits long). The status of a job can be checked by typing the following command: qstat xxxxxx, where xxxxxx is the job number. The server also generates an error file, which resides in the root directory of the home scratch space. If a job is run successfully, the size of this file should be 0kb. After the job is run successfully, the output can be viewed by looking at the .out file specified in the .sh script.

Considering the number of genes in common between each pair of genomes, we can confirm the phylogenetic relationships of these taxa. We can say that, MR4 and MR7 are more closely related and MR1 is more divergent. This agrees with the fact that MR4 and MR7 come from the same ecological niche, the black see, and MR1 comes from Oneida Lake (New York). In addition, a survey of the 68 core gene set (Table 4) demonstrates that most members of this list are housekeeping genes. This is consistent with our original hypothesis, for when species diverge, they adapt to their own environment. Also consistent with the original hypothesis is that non-core genes are related to survival in different environments and conditions. One example of housekeeping genes observed in this core gene set is ribosomal proteins; since they have a vital role in the physiology of the cell, they should be highly conserved.

## Discussions and Conclusions

It must be stressed that this method underestimates the number of conserved genes. For instance, consider the gene lists for species A and B. In a given Blast search, the best match to gene A1 is gene B2, while the reciprocal blast search yields a match between B2 and gene A2. For this example, it may be that this transitive relationship is comparable to a 1:1 relationship in terms of relatedness. Another issue is that our criteria yielded no multiple best matches for a given gene in the list. This type of result allows us to ignore genes that diverged after gene duplication event duplication, but does not do so by considering their function. The "core gene" methodology does not preserve results such as these. Thus, the core genome concept is a bit abstract as it relates to function. However, we feel that this is the best way to get at fundamental evolutionary relationships.

We conclude that this is a good method to identify a core set of genes among a group of species. In the future, it would be interesting to test it with a larger set of species, been a larger set of bacterial species or even using eukaryotic taxa to compare how much of the genome has being conserved across significant periods of evolution. One potential if not ambitious future application would be to compare whole-genome databases from humans, chimps, and macaques similar to the "genomic triangulation" method of Harris et al. (2007). This would be a good complement to papers which make inferences regarding interspecific evolutionary differences by looking for shared and unique positively selected genes (Bakewell et al. 2007).

**Appendix:**

Procedure for installing and running Blast on an HPCC cluster: Install Blast from the following archive: ftp://ncbi.nih.gov/bin/blast/executables/LATEST/blast-2.2.16-ia32-linux.tar.gz. Extract the archive to an install directory on the home scratch space. This should install the program. The script must be submitted to one of the clusters (type either "ssh xxxx" or "ssh xxxx", where xxxx is the name of the specific cluster, at the command line to submit a job). If the job is not properly submitted to one of the clusters, the Blast .exe files (formatdb, blastall) will not compile upon submission of the job.

**Figures and Tables:**

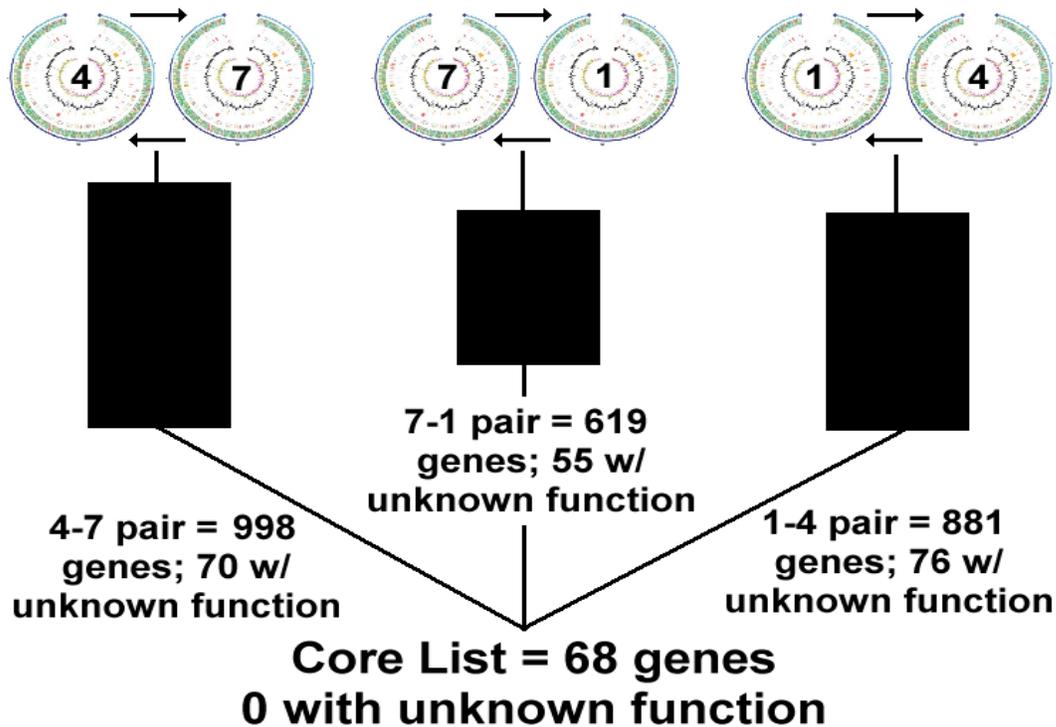

Figure 1: procedure for reciprocal Blast search in graphical form.

**Table 1: Script for submitting the jobs to HPC (in this case, an all genes-all genes search of MR-4 and MR-7 (For each type of biological data, different programs internal to Blast are used. For DNA sequences, blastn was used; for protein sequences, blastp was used).**

```
#!/bin/sh
#PBS -M aliceabr@msu.edu -m abe
#PBS -l nodes=1:ppn=1,walltime=8:0:0,mem=1gb
~/blast-2.2.16/bin/blastall -p blastp -d ~/project/input-output/genes/
MR4_finished.fasta -i ~/project/input-output/genes/MR7_finished.fasta -o
~/project/input-output/genes/pair4_7.out -F F -e 1E-10 -v 1 -b 3 -X 150 -q -1 -a 2 -m 9
```

**Table 2: Code for formatting proteins sequences into a searchable input file (NOTE: formatting proteins are different than formatting DNA sequences).**

```
#!/bin/sh
#PBS -M aliceabr@msu.edu -m abe
#PBS -l nodes=1:ppn=1,walltime=8:0:0,mem=1gb
~/blast-2.2.16/bin/formatdb -i ~/project/input-output/MR7_finished.fasta -p T -n
~/project/input-output/MR_7_finished
```

**Table 3: Python Script for sorting results by an identity threshold (a variable value for the threshold identity can be passed in).**

```
def criterion("file", ID):
  inp = open("file")
  inl = inp.readlines()
  sim = ID
  qs = ""
  for i in inl:
     L = i.split("\t")
     s = (L[2])
     if s > sim:
        sim == s
        qs == "%s-%s"
  print L[0],L[1],qs,L[3],L[4],L[5],L[6],L[7],L[8],L[9],L[10],L[11]

criterion("blast.out", 90)
```

**Table 4: List of core gene set with gene annotations.**

| | |
|---|---|
| 1 | SirA family protein 16858:17103 r |
| 2 | carbonic anhydrase, family 3 4033 |
| 3 | transcriptional regulator, BadM/R |
| 4 | NLP/P60 protein 70622:71101 forwa |
| 5 | transcriptional regulator, GntR f |
| 6 | Twin-arginine translocation pathw |
| 7 | 3,4-dihydroxy-2-butanone 4-phosph |
| 8 | Methyltransferase type 11 157051 |
| 9 | 3'(2'),5'-bisphosphate nucleotida |
| 10 | transcriptional regulator, TetR f |
| 11 | ribosomal protein S12 220347:2207 |
| 12 | ribosomal protein S7 220802:22127 |
| 13 | ribosomal protein S19 227800:2280 |
| 14 | ribosomal protein S8 231755:23214 |
| 15 | ribosomal protein L36 235565:2356 |
| 16 | ribosomal protein S13 235794:2361 |
| 17 | ribosomal protein S4 236589:23720 |
| 18 | ribosomal protein L17 238264:2386 |
| 19 | CcmE/CycJ protein 239841:240326 r |
| 20 | Heme exporter protein D (CcmD) 24 |
| 21 | cytochrome c, class I 242945:2432 |
| 22 | periplasmic protein thiol--disulp |
| 23 | Redoxin domain protein 247999:248 |
| 24 | acetylglutamate kinase 260668:261 |
| 25 | ABC transporter related 1050081:1 |
| 26 | LrgB family protein 1054867:10556 |
| 27 | peptidylprolyl isomerase, FKBP-ty |
| 28 | transcriptional regulator, AraC f |
| 29 | NADH:ubiquinone oxidoreductase, s |
| 30 | Ferritin, Dps family protein 1212 |
| 31 | purine nucleoside phosphorylase 1 |
| 32 | phosphoserine phosphatase SerB 12 |
| 33 | thioesterase superfamily protein |
| 34 | conserved hypothetical protein 13 |
| 35 | MazG family protein 1384031:13849 |
| 36 | stationary-phase survival protein |
| 37 | carbon storage regulator, CsrA 14 |
| 38 | DNA polymerase III chi subunit, H |
| 39 | cytochrome c, class II 1411184:14 |
| 40 | homoserine kinase 1416481:1417419 |
| 41 | sigma 54 modulation protein/ribos |
| 42 | Fe(II) trafficking protein YggX 1 |
| 43 | Glutaminase 1462192:1463106 forwa |
| 44 | transcriptional regulator, LysR f |
| 45 | non-canonical purine NTP pyrophos |
| 46 | NrfJ-related protein 1493842:1494 |
| 47 | Pirin domain protein domain prote |
| 48 | transcriptional regulator, LysR f |
| 49 | conserved hypothetical protein 18 |
| 50 | two component transcriptional reg |
| 51 | hypothetical protein 1917957:1918 |
| 52 | multiple antibiotic resistance (M |
| 53 | protein of unknown function DUF86 |
| 54 | ATPase associated with various ce |
| 55 | DsrE family protein 2369396:23697 |
| 56 | CrcB protein 2371879:2372253 reve |
| 57 | outer membrane lipoprotein carrie |
| 58 | transcriptional regulator, AsnC f |
| 59 | thioredoxin reductase 2379672:238 |
| 60 | ribosomal protein L20 2380810:238 |
| 61 | integration host factor, beta sub |
| 62 | NUDIX hydrolase 2607606:2608193 f |
| 63 | conserved hypothetical protein 26 |
| 64 | 2OG-Fe(II) oxygenase 2681404:2682 |
| 65 | nicotinamide mononucleotide trans |
| 66 | conserved hypothetical protein 28 |
| 67 | conserved hypothetical protein 28 |
| 68 | Patatin 2969561:2970508 reverse M |

**References:**

**Bakewell, M.A., P. Shi, and J. Zhang.** 2007. More genes underwent positive selection in chimpanzee evolution than in human evolution. Proc. Natl. Acad. Sci. USA **104(18):** 7489–7494.

**Bejerano, G., M. Pheasant, I. Makunin, S. Stephen, W.J. Kent, J.S. Mattick, and D. Haussler.** 2004. Ultraconserved elements in the human genome. Science **304(5675):** 1321-1325.

**Cicarelli, F.D., T. Doerks, C. von Mering, C.J. Creevey, B. Snel, and P. Bork.** 2006. Toward automatic reconstruction of a highly resolved tree of life. Science **331:**1283-1287.

**Daubin, V., N.A. Moran, and H. Ochman.** 2003. Phylogenetics and the cohesion of bacterial genomes. Science **301:**829-832.

**Fuschman, C.A. and G. Rocap.** 2006. Whole-genome reciprocal BLAST analysis reveal that *Planctomycetes* do not share an unsually large of genes with *Eukarya* and *Archaea.* Appl. Enviro. Microbiol. **72:**6841-6844.

**Harris, R.A., J. Rogers, and A. Milosavljevic.** 2007. Human-Specific Changes of Genome Structure Detected by Genomic Triangulation. Science **316(5822):**235-237.

**Konstantinidis, K.T. and J.M. Tiedje.** 2005a. Genomic insights the advance the species definition for prokaryotes. Proc. Natl. Acad. Sci. USA. **102:**2567-2572.

**Konstantinidis, K.T. and J.M. Tiedje.** 2005b. Towards a genome-based taxonomy of prokaryotes. J. Bact. **187:**6258-6264.

Mira, A., H. Ochman, and N.A. Moran (2001). Deletional bias and the evolution of bacterial genomes. Trends Genet. **17:**589-596.

**Nierman, W.C., R.L. Strausberg, C.M. Fraser, S. Zhao, J. Shetty, L. Hou, A. Delcher, B. Zhu, K. Osoegawa, and P. de Jong.** 2004. Human, Mouse, and Rat Genome Large-Scale Rearrangements: Stability Versus Speciation. Genome Res. **14:** 1851-1860.

**Overbeek, R., T. Begley, R.M. Butler, J.V. Choudhuri, H-Y. Chuang, M. Cohoon, et al.** 2005. The subsystems approach to genome annotation and its use in the project to annotate 1000 genomes. Nucl. Acid. Res. **33:**5691-5702.

**Wei, L., Y. Liuc, I. Dubchakd, J. Shona, and J. Park.** 2002. Comparative genomics approaches to study organism similarities and differences. J. of Biomed. Inform. **35(2):** 142-150.